# SYSTEM OF BIBLIOMETRIC MONITORING SCIENCES IN UKRAINE

*Kostenko Leonid, Zhabin Alexander, Kuznetsov Alexander,*
*Kukharchuk Elizabeth, Simonenko Tatiana*

Vernadsky National Library of Ukraine
Holosiivskyi ave., 3, Kiev, Ukraine

## Abstract

The origins of scientometrics (research metrics) were analysed and the lack of attention to elaboration of its methodology was emphasized. The approaches to evaluation of scientific activity outcome were considered and the tendency of transition from formal quantitative indicators to receiving expert conclusion on the basis of bibliometric indicators was noted. The principles of the Leiden manifesto of scientometrics were set out, keeping to which provides clear monitoring and support of science development, and favours establishing of the constructive dialog between scientific environment and society as well. The conceptual statements and peculiarities of practical realization of the informative and analytical system "Bibliometryka Ukraynskoyi Nauky" ("Bibliometrics of the Ukrainian Science") elaborated by the Vernadsky National Library of Ukraine, are being represented. The proposals on the formation of advisory councils, which are aimed to adopt conclusions on the effectiveness of research activity of institutions, were considered. The feasibility of building a common platform for expert evaluation of scientific studies for countries of the Eastern Partnership by initiating similar bibliometric projects in these countries and their further convergence is proved.

## Introduction

Term "scientometrics" was put into scientific circulation by V. Nalimov in 1969. In his paper [1] it was suggested: «Let us give the term "scientometrics" to qualitative methods of scientific studying and development as information process». The issues of scientometric study include the following questions: information model of scientific development, growth of informative flows, literature citing, studying of internal connections in science according to the language of bibliographic references, evaluation of contribution brought into world scientific information flow by separate countries, process of statistical analysis of scientific trends. Positively evaluating Nalimov's contribution to scientometrics, we should note the negative role of his interpretation of this term since such a definition had been oriented further researches in this sphere at "numerological" way of development.

Ukrainian scientists have long-term experience in theoretical and practical developments. Special meaning for their evolution had a fundamental monograph of the founder of the Center of research and technical development potential and history of science (NAS of Ukraine) Dobrov G. "Research about research: introduction to the general knowledge about science", which started the beginning of this trend of works. The monograph deepened an interest to scientific researches in general and was subsequently republished in many countries of the world [2]. In the monograph he made an emphasis on the need for a systematic study of trends and perspectives of science development in Ukraine and in the world as well. It is reflected in a wide spectrum of the considered questions such as history of the development of science and scientific schools, conditions and trends of scientific and technological potential, infrastructure of science, scientific and technological policy, innovation policy, international elaboration issues etcetera. The definition of study about science offered by Dobrov "… is a complex research and theoretical generalization of the experience of social systems' functioning in science with the aim to ground scientific and technical policy and also rational forming of scientific potential and increasing of research activity effectiveness by means of social, economical and organizational influence" which is topical even nowadays. It reflects systemic structure of researches about science and the need for complex knowledge about science. It follows from this definition that scientometrics should be based on theoretical and methodological developments of study about science.

Unfortunately, Dobrov's ideas as for the need to direct scientometrics at the support of research didn't receive proper dissemination. As an exception one can list the works of Korennoi A. [3], Marshakova I. [4], Haytun S. [5]. The prior attention in these works was paid to organization of scientific prognostication, use of bibliometric indicators for determining the structure of science and monitoring its development, as well as to the shortcomings of the use of quantitative indicators by assessing the effectiveness of scientific research. We should emphasize that Marshakova I., together with Haytun S. considered the definition of scientometrics, given by Nalimov V, "too much categorical" [6].

In practical aspect, the greatest contribution to scientometric study was made by J. Garfield. He offered a unique idea on the use of scientific references as a means of information retrieval and study of the structure of science. His name is associated with the organization of the Institute for Scientific Information US and creating a database Web of Science with the analytical add-ons. At the same time, Garfield Y. himself tirelessly called for caution in the use of citation data noting that they, "as any instrument – from nuclear energy to the hammer – must be properly employed" [7].

The omission of Garfield's warning, ignoring such task of scientometrics as science of science and straightforward focus on "numerology" by V. Nalimov led to the emergence of scientific methods for assessing the effectiveness of research activity which did not sufficiently take into account considerable aspects of scientific and research work, being a combination of various kinds of formal parameters [8-10].

Currently dominates such a point of view that only professional expertise can provide a comprehensive and objective assessment of scientific results; the bibliometric indicators, in their turn, serve as a supportive tool for taking a decision by experts [11-14].

The need for an objective evaluation of the effectiveness of research and exceptions of lobbying for certain scientometric databases requires consolidated information from different sources.

## Purpose of the article

The purpose of this paper is developing a theoretical framework for the creation of a common platform for monitoring and evaluation of the expert support of science and forecasting its development.

# Methodology

Before defining the basic concepts of the construction of this platform, it is useful to consider the latest developments in the methodology of evaluation of scientific activity. In a concentrated form they are set out in the Leiden Manifesto for Scientometrics, adopted at the 19th International Conference of Science and Technology Indicators «Context Counts: Pathways to Master Big and Little Date» (3-5 September 2014, Leiden, The Netherlands) and published in the journal "Nature" in April 2015 [15]. Its ten principles are nothing new for specialists engaged in bibliometrics and scientometrics, although previously they were not set out in a systematic way.

1. Quantitative evaluation should support qualitative, expert assessment.

2. Measure performance against the research missions of the institution, group or researcher.

3. Protect excellence in local relevant research.

4. Keep data collection and analytical processes open, transparent and simple.

5. Allow those evaluated to verify data and analysis.

6. Account for variation by field in publication and citation practices.

7. Base assessment of individual researches on a qualitative judgment of their portfolio.

8. Avoid misplaced concreteness and false precision.

9. Recognize the systemic effects of assessment and indicators.

10. Scrutinize indicators regularly and update them.

The first principle is fundamental and presupposes primacy of expert evaluation over "numerological" – formal indicators should be collected and taken into account when assessing, but only as a part of the information required for a professional expert analysis. This principle is closely related to the seventh, which recommends to take into account portfolio of a scientist or a team (experience, achievements, authority). One should draw his attention to the third principle, which speaks about the importance of specific indicators to assess regional studies that are of national importance and published in the non-English lanaguage journals (as an example for Ukraine one can point to the environmental monitoring of the Chernobyl zone). An important principle is openness of data and procedures of analysis, which is not always provided by commercial bibliometric systems. It is necessary to mark the ninth principle too, which warns about danger of an assessment on a single indicator, leading to a game with him and to substitution of research objectives - the aim is to reach the maximum of this indicator. If this factor relates to the Hirsch index, the task of the scientist may be aimed at "cheating" but not at discovery of new laws and identifying previously unknown patterns.

From the analysis of the principles of the Leiden Manifesto of Scientometrics implies that scientometrics should not focus on the support of administrative reform processes of education and science, but to promote their development, particularly in a search for breakthrough research front, that is, its purpose – to support decisions of not "political", but scientific problems.

## Tools

We have developed the information and analytical system "Bibliometrics of the Ukrainian Science" which correlates to the Principles of the Leiden Manifesto of Scientometrics [16]. The system is:

- Register of the Ukrainian scientists who have created their bibliometric profiles in Google Scolar;
- Single window access to bibliometric indicators of scientists, groups and magazines in the leading science-metric systems (Scopus, Web of Science, Russian Science Citation Index, Ranking Web of Research Centers);
- Analytical processing tools of bibliometric data for receiving the information about the industry, departmental and regional structure of the Ukrainian science;
- Source base for expert assessment of scientific activity and identifying of new trends in the development of science;

- National component of the project Ranking of Scientists (Cybermetrics Lab).

A speciality of the system is that it was designed with the concepts of the convergence of international and national bibliometric projects. These concepts stipulate a basic consolidation platform of bibliometric data from various systems, a common system of categories and subcategories (the classification scheme) for representing the areas of knowledge and tools of analytical calculations for expert evaluation and the identification of trends in the development of science.

The main criteria for choosing a consolidation platform of bibliometric data were its accessibility and scope of the indexed scientific papers for obtaining reliable results in statistical terms. Today, the given conditions best meet bibliometric platform Google Scholar, which handles the entire global scientific documentary flow excluding the materials with limited access. Peer-reviewed papers, dissertations, books, abstracts, conference proceedings and other scientific literature from different fields of research are being indexed. The abovementioned positive qualities of Google Scholar have been evaluated by a number of institutions.

In particular, the research team Cybermetrics Lab (Spain) chose it as a base platform for scientists rating based on their public bibliometric profiles [17]. Considering the fact that Google Scholar is the starting point for information search, the owners of commercial scientometric systems are making efforts for the organization of mutually beneficial cooperation with it. So, on the official website of the Thomson Reuters Corporation there is information on collaboration with Google Scholar [18]. It can be assumed that the Elsevier Corporation will follow the same path either. In this case, the use of Google Scholar will allow to receive data from the abovementioned commercial systems in the presence of a license to access them.

The service "Bibliographic references" by Google Scholar provides the ability to create bibliometric profiles that can be viewed as a portfolio of scientists and collectives. They contain information on the ordered lists of their works, chart citation in the time slice, and information about the affiliation of organizations and magazines as well. This service is in demand - as of September 2015 it was used by more than 10 thousand researches from the Ukrainian segment of the Internet. Among them are world famous scientists and beginning researchers with several publications. Such an amount provides a first glimpse of the intellectual potential of the country, reflecting its regional, departmental and thematic sections.

Particular attention should be paid to a uniform system of categories and subcategories (classification schemes, subject headings) for the submission about scientific disciplines. In the library and information practice, the greatest application has Universal Decimal Classification. But it focuses on the substantive assessment of a separate document (book, article) and not on the particular field of study in which the researcher is working. However, classifiers of scientific specialties are free from this shortcoming at defenses of dissertations. But they are inappropriate to be applied in integrational bibliometric projects due to the lack of harmonization between classifications of different countries.

Appropriate solutions to the representation of disciplines are the categories and subcategories, offered by the leading scientific and information corporations, among which we should highlight Google Scholar, Elsevier and Thomson Reuters. Each of them offers its own classification system, which is a collection of about 300 categories and sub-categories, which are determined on the basis of processing of English documentary flows and harmonized with modern concepts and categorical apparatus of science. Taking into account choice of base platform Google Scholar for consolidation of bibliometric data, it seems appropriate to use categories and subcategories to represent the branches of knowledge [19].

The principal difference of bibliometric systems from bibliographic databases and, in particular, from the electronic catalogs is the presence of tools of analytical calculations to support the expert assessment and to identify trends in the development of science. In the system Web of Science the function of such a tool belongs to a superstructure InCites, which provides an opportunity to assess and compare the results of scientific research of organizations and countries to define their place in the world of science. Such a superstructure SciVal by the Elsevier Corporation is based on resource database Scopus. It helps organizations to assess their potential and identify promising development strategy. Based on the analysis of co-citation and visualization technique, this superstructure creates a unique graphic card or "Wheel of Science" which illustrates efficiency of the organization in all scientific disciplines. InCites and SciVal are useful for the analysis of scientific activity of a separate organization, region or a country. The choice depends on the goal: for strategic planning of scientific activity in organization and choice of support directions, the SciVal by Elsevier Corporation should be used; and for comparison with other specific organizations or monitoring the activity of individual scientists, research groups and branches of science – InCites by Thomson Reuters Corporation [20] .

The abovementioned analytical superstructures have been perfected over a long period of time and have great functionality. Analytical calculation tools of "Bibliometrics of the Ukrainian Science" in terms of functionality, is inferior to InCites and SciVal. Nevertheless, it allows you to get a general idea about the state of the Ukrainian science and its sectoral, departmental and regional distribution. Indicators of sectoral distribution evidence about the prevalence of specialists in economics - they make up about 25% of the total number of the Ukrainian scientists submitted in Google Scholar. In the departmental aspect dominates scientific and teaching staff of the Ministry of Education and Research (60%); in the regional - scientists from Kiev (35%). Among highly cited researchers (with Hirsch index more than 25) is the majority of employees of the National Academy of Sciences (65%) [21].

Works on the improvement of this analytical apparatus are continuing, in particular in terms of formation and use of linguistic ontology as a means to identifying trends in the development of science. The information base for creating this ontology acts as bibliometric profiles of scientists who provide the verified data on their writings. An analysis of frequency words indicators from the titles of publications within one subcategory Google Scholar allows you to select the most frequently used scientific terms and to identify trends in the basic science by comparing terminology systems through different years. At the same time with the help of the frequency dictionary of new words one can carry out an expert forecasting of the development of science and to identify original articles that deserve special attention [22].

A database management system MySQL was used as a basic software of the "Bibliometrics of the Ukrainian Science" This system satisfies the requirements of the so called "cross platform", free distribution, open source code and integrity with programming languages such as java, perl, php, python.

**Performance evaluation**

It should be emphasized that quantitative indicators of the "Bibliometrics of the Ukrainian Science" can not be considered as criteria for assessing the effectiveness of research activities. They are a source base for the adoption of expert solutions. Evaluation of several hundred scientific organizations require a large number of experts, and very significantly, that they enjoy the confidence of colleagues just as worthy of scientific experts in their field. Therefore, the full body of experts should be formed by referring to academic councils (in Science and Technology field) of all of certified institutions with suggestion to nominate experts for each of the scientific directions of the organization, and to provide the necessary range of professional information about each nominee.

Comparative assessment of scientific impact is advisable to carry out within the so-called reference groups of research organizations which need to be formed on the basis of their proximity to areas of their scientific activity and the types of the results (basic research, technological developments, scientific and technical services, and so on.). For each reference group should be formed its advisory council. The brunt of the work falls on these tips. The total control over the process and approval (or correction request) of the results of the tips should be entrusted to a single commission for the evaluation of performance. In the case of the low performance of formal indicators of organization, its disagreement with the assessment as a whole, more detailed assessment must be implemented, including the examination of each unit [23].

### Perspectives

The positive experience of testing the developed conceptual positions in the operation of information and analytical system "Bibliometrics of the Ukrainian Science", for 2014-2018, showed their validity and applicability for implementing bibliometric projects focused on the subsequent convergence. They in particular can be used for the initiation of the project "Bibliometrics of Science of Eastern Countries Partnership".

In its framework a Member State assumes responsibility for the creation of the English-language database with information on bibliometric profiles of its scientists in the Google Scholar System. The content of the database is transferred to the integration center, which will be defined in the deployment of work. This center handles national bibliometric segments and creates corporative resource that will be available to all the project participants. Moreover, the integration center supports in a free access consolidated citation information with analytical tools for obtaining information about the contribution of each country into the system of scientific communications, regional and sectoral distribution of researchers and research groups, their formal and informal relationships, in a free access.

The advantage of the proposed project is, above all, the possibility of obtaining a single bibliometric database for comparison and expert evaluation of scientific activity in the countries of the Eastern Partnership. No less important is the fact that the project will contribute to strengthening the linkages between researchers and improve the positive image of science.

### Conclusions

1. The original definition of scientometrics as a set of quantitative methods of analysis and assessment of science for a long period predetermined "numerological" path of its development. Development of theoretical basis of scientometrics led to a new understanding of this term. Today, it is a tool for monitoring and expert support to the development of science.

2. The essence of the modern methodology of evaluating the effectiveness of research in a concentrated form is set out in the ten principles of "The Leiden Manifesto of Scientometrics" which target it on the transparent monitoring of the scientific sphere for subsequent expert evaluation.

3. The information and analytical system "Bibliometrics of The Ukrainian Science", which was developed by our team, has become the national component of the project Rankings of Scientists (Spain) and complies with the principles of The Leiden Manifesto of Scientometrics.

4. Building a common platform for expert evaluation of scientific research of the Eastern Countries Partnership can be achieved by initiating similar bibliometric projects in these countries and their subsequent convergence. The implementation of the program is possible with grant support.